\documentclass[12pt]{iopart}

\usepackage{color}
\usepackage{epsf}
\usepackage{graphicx}
\usepackage{iopams}  
\begin{document}


\title[Estimate of the Coulomb Correlation Energy in CeAg$_2$Ge$_2$] {Estimate of the Coulomb Correlation Energy in CeAg$_2$Ge$_2$ from Inverse Photoemission and High Resolution Photoemission Spectroscopy} 

\author{Soma Banik$^{1,*}$, A. Arya$^2$, Azzedine Bendounan$^3$, M. Maniraj$^4$, A. Thamizhavel$^5$, I. Vobornik$^6$, S. K. Dhar$^5$, S. K. Deb$^1$}
\address{$^1$Indus Synchrotron Utilization Division, Raja Ramanna Centre for Advanced Technology, Indore, 452013, India.}
\address{$^2$Materials Science Division, Bhabha Atomic Research Centre, Mumbai 400085, India.}
\address{$^3$Synchrotron SOLEIL, L$'$Orme des Merisiers, Saint-Aubin, BP 48, FR-91192 Gif-sur-Yvette Cedex, France.}
\address{$^4$UGC-DAE Consortium for Scientific Research, Khandwa Road, Indore, 452001,India.}
\address{$^5$Tata Institute of Fundamental Research, Homi Bhabha Road, Colaba, Mumbai 400005, India.}
\address{$^6$Istituto Officina dei Materiali (IOM)–CNR, Laboratorio TASC, Area Science Park, S.S.14, Km 163.5, I-34149 Trieste, Italy.}
\ead{$^*$soma@rrcat.gov.in}

\begin{abstract} 
The occupied and the unoccupied electronic structure of CeAg$_2$Ge$_2$ single crystal has been studied using high resolution photoemission and inverse photoemission spectroscopy respectively. High resolution photoemission reveals the clear signature of Ce $4f$ states in the occupied electronic structure which was not observed earlier due to the poor resolution. The coulomb correlation energy in this system has been determined experimentally from the position of the $4f$ states above and below the Fermi level. Theoretically the correlation energy has been determined by using the first principles density functional calculations within the generalized gradient approximations taking into account the strong intra-atomic (on-site) interaction Hubbard $U_{eff}$ term. Although the valence band calculated with different $U_{eff}$ does not show significant difference, but the substantial changes are observed in the conduction band. The estimated value of correlation energy from both the theory and the experiment is $\approx$4.2~eV for CeAg$_2$Ge$_2$.  
\end{abstract}


\maketitle

\section{Introduction}
 Unusual properties in rare earth based systems can be understood on the basis of the strong Coulomb interaction within the $4f$ shell and the hybridization between the partially delocalized $f$ electrons and the non-$f$-band electrons \cite{Sheng, Kang, Lucas}. The presence of strong electron-electron correlation in Cerium metal and its compounds is well known \cite{Lang, Goraus12, Antonov11, Konishi00, Hufner2}. For the magnetic systems which shows variety of spin ordering, the electron-electron correlation has been reported to play an important role \cite{Goraus12, Antonov11, Konishi00, Soma, Si01}. CeAg$_2$Ge$_2$ shows the anisotropic magnetic properties with antiferromagnetic transition at 4.6~K \cite{Thamizhavel08}. The antiferromagnetic ground state in this system is due to the competition between the Ruderman-Kittel-Kasuya-Yosida (RKKY) interaction and the Kondo effect which is mainly due to the presence of Coulomb interaction \cite{Thamizhavel08, Lubritto95}. Hence, it is important to determine the Coulomb interaction in this system experimentally. 
 
 Recently we have carried out a detail study of the electronic structure of CeAg$_2$Ge$_2$ by resonant photoemission (RPES) and angle resolved photoemission spectroscopy (ARPES) \cite{Soma1,Soma2}. RPES showed two resonance features in the valence band spectra related to Ce $4f$ states. Ce is found to be in the trivalent state in CeAg$_2$Ge$_2$ from the X-ray photoemission measurement \cite{Soma1}.  ARPES on the CeAg$_2$Ge$_2$ (001) surface along the $\Gamma$-Z direction revealed that the Ce 4f band which is near to the Fermi level (E$_F$) shows a clear dispersion while the Ce 4f band at higher binding energy doesn$'$t show a dispersion \cite{Soma2}. The occupied electronic structure of CeAg$_2$Ge$_2$ could very well be explained from the theoretical calculation without considering the electron-electron correlation term \cite{Soma1,Soma2}. However, for Ce metal the effect of correlation has been reported to be more evident in the unoccupied part of the electronic structure than the occupied part \cite{Lang, Herbst78}. Hence, it is of interest to study the unoccupied part of the electronic structure of CeAg$_2$Ge$_2$ by inverse photoemission spectroscopy (IPES) to extract the information about the correlation in this system. Moreover, in our earlier photoemission studies, due to the poor experimental resolution (400~meV) the Ce $4f$ feature near the E$_F$ was not observed very clearly \cite{Soma1} and the correlation energy could not be estimated. So, there is a need to do the high resolution photoemission (PES) measurement on this system. Therefore, in the present work the electronic structure of CeAg$_2$Ge$_2$ has been studied using high-resolution PES and IPES. The electron correlation energy ($U$) has been determined experimentally from the PES and IPES spectra. GGA+U calculations have been performed to determine $U_{eff}$ theoretically and to compare it with the experimental results.

\section{Methods}
CeAg$_2$Ge$_2$ single crystal was grown by the self-flux method \cite{Thamizhavel08}. High resolution photoemission measurements at 20~meV resolution were performed at the high-resolution photoelectron spectroscopy station of TEMPO beamline at Synchrotron SOLEIL, France \cite{Polack10} and also at APE beamline of Synchrotron Elettra, Italy \cite{Panaccione}. The clean surface of the CeAg$_2$Ge$_2$ single crystal was obtained by cleaving the sample in-situ in a base pressure of 9 $\times$ 10$^{-11}$ mbar and at a temperature of about 40~K.  The data have been recorded with a Scienta SES 2002 electron energy analyzer. The photoemission measurements with 400~meV resolution was carried out at the angle-integrated PES beamline on the Indus-1 synchrotron radiation source, India \cite{Chaudhary02}. The binding energy in the photoemission spectra has been determined with reference to the Fermi level of the clean Gold surface that is in electrical contact with the sample at the same experimental conditions\cite{Soma1}. IPES experiments were performed under ultrahigh vacuum at a base pressure of 2$\times$10$^{-10}$~mbar in a separate chamber. To obtain atomically clean surface the sample has been sputtered with 1.5~keV argon ions and annealed at 500~K \cite{Soma1}. An electrostatically focused electron gun of Stoffel Johnson design and an acetone gas filled photon detector with a CaF$_2$ window have been used for the IPES experiments \cite{Soma3}. IPES has been performed in the isochromat mode where the kinetic energy of the incident electrons has been varied at 0.05~eV steps and photons of fixed energy (9.9~eV) are detected with an overall resolution of 0.55~eV \cite{Soma3}. Similar to PES study, in IPES also the binding energy has been determined with reference to the Fermi level of the clean Silver surface that is in electrical contact with the sample at the same experimental conditions. In both PES and IPES experiments, the incident angle and the take off angle are kept fixed at 45 degree. 

The spin-polarized electronic structure calculations have been performed using the density-functional theory (DFT)\cite{Hohenberg65} and a very accurate full-potential linear augmented plane-wave (FP-LAPW) approach incorporating the spin-orbit (SO) coupling as implemented in WIEN2K code \cite{Blaha01}. This is an implementation of a FP-LAPW plus local orbitals (LAPW + lo) method \cite{Singh06} within the DFT. The Perdew, Burke, Ernzerhof (PBE) \cite{Perdew96} gradient corrected local spin density approximation (LSDA-GGA) for the exchange correlation (XC) potential was used. The spin-orbit (SO) interaction was treated by the second-variational approach \cite{Singh06,Sjostedt00}. A plane wave expansion with R$_{MT}$ $\times$ K$_{max}$ equal to 8 and the dependence of the total energy on the number of $k$ points in the irreducible edge of the first Brillouin zone had been explored within the linearized tetrahedron scheme by performing the calculation for 405 $k$ points (17$\times$17$\times$17 mesh). The cut-off for charge density was fixed at $G_{max}$ = 14. The muffin-tin radii used for the calculations were 2.9, 2.4 and 2.1 Bohrs for Ce, Ag and Ge, respectively. To account for the Coulomb correlation interaction within the Ce-$4f$ shell, we additionally considered the PBE XC potential corrected according to the GGA+U method \cite{Tran06, Larson07}. For Ce, the values of $U_{eff}$ (=$U-J$ with $J$= 0.95~eV) were varied in turn by taking it to be 1.4~eV, 2.7~eV, 3.4~eV, 4~eV and 5.4~eV respectively. In the calculation $U$ and $J$ do not enter separately; hence the difference $U_{eff}= U-J$ is meaningful \cite{Dudarev} and has been compared with the experimental $U$.
 To determine the ground state of the system we have adopted a standard procedure of minimizing the total energy as a function of lattice parameters ($a$ and $c/a$). We have also performed the atomic position optimization such that the residual force on each atom was less than 1 meV/$\AA$. The  equilibrium lattice parameter $a$ was calculated to be 4.313 $\AA$ with $c/a$= 2.572 which agrees well with the experimental value of $a$= 4.301  $\AA$ (and $c/a$= 2.551) \cite{Thamizhavel08}.  The relaxed ionic positions for Ce: (0,0,0); Ag: (0,0.5,0.25); and Ge:(0,0,0.391) closely match experimental values of Ce: (0,0,0); Ag: (0,0.5,0.25); and Ge:(0,0,0.389) \cite{Thamizhavel08}.

\section{Results and discussions}


Basically the electron correlation energy $U$ is the energy required to transfer an electron within the same band between two ions in a solid which are initially in the same valence state \cite{Hufner1, Hufner2}. For Ce metal and its compounds in the trivalent state, $U$ is the energy to produce an Ce$^{2+}$ and an Ce$^{4+}$ ion by transferring an electron from one Ce$^{3+}$ ion to another.  Hence, $U$ can be derived experimentally from the position of the $4f$ states above and below E$_F$ from the valence and conduction band using PES and IPES respectively \cite{Anisimov, Anisimov2, Lang, Baer}.

In Figure 1 we have shown the influence of resolution on the PES spectra for the experiments carried out at two different energy resolutions of 20~meV and 400~meV. Generally for Ce metal and its compounds the peak structure pinned at E$_F$ corresponds to $f^1$$_{5/2}$ final state, whereas the peak at 280~meV  corresponds to $f^1$$_{7/2}$ final state \cite{Patthey}. The intensity ratio of $f^1$$_{5/2}$/$f^1$$_{7/2}$ is related to the hybridization strength as shown for Ce $4f$ states in CeNi$_2$Al$_5$ and CeNiAl$_4$ \cite{Kashikuara}. In Ce alloys 4$f^1$$_{7/2}$ and 4$f^1$$_{5/2}$ features can appear due to magnetic Ce $4f$ states as well as due to the Kondo singlets formed via quantum entanglement of the $4f$ states with the conduction electron states \cite{Patthey, Kalo}. The appearance of 4$f^1$$_{7/2}$ and 4$f^1$$_{5/2}$ feature in CeAg$_2$Ge$_2$ is mainly due to the magnetic nature of Ce in this system\cite{Thamizhavel08}. Figure 1 shows a clear difference in the two spectra for the 4$f^1$$_{7/2}$ excitation (marked by tick in Figure 1) which is more prominent in the high resolution 20~meV spectra. Hence, the value of  spin orbit splitting can be obtained from the high resolution spectra, as shown in the inset of Figure 1. The spin orbit splitting ($\Delta_{S.O.}$) between 4$f^1$$_{7/2}$ and 4$f^1$$_{5/2}$ is found to be 280 meV which is in good agreement with the other Ce based systems \cite{Patthey}. In the inset of Figure 1 we have compared the on-resonance spectra at h$\nu$= 121 eV with the off-resonance spectra at h$\nu$= 80 eV for the PES measurement carried out at high resolution. There is a negligible signature of the Ce 4$f^1$$_{7/2}$ feature in the off-resonance spectra. This clearly signifies the importance of RPES, to identify the position of the Ce $4f$ features in the valence band. The shapes of the 4$f^1$$_{7/2}$ and 4$f^1$$_{5/2}$ features indicates that Ce in this system is $\gamma$-type \cite{Patthey}. The 4$f^1$$_{7/2}$ excitation in the spectra with 400~meV resolution is not clearly distinguishable because the resolution is more than the spin orbit splitting. The effect of the instrumental resolution in determining the Ce $4f$ features has also been shown for CeAl$_2$ alloy \cite{Patthey, Patthey2, Parks}. Hence Figure 1 clearly illustrates the necessity of high energy resolution and the resonance excitation energy in order to extract the reliable information of Ce $4f$ features near E$_F$ in the PES spectra \cite{Hillbrecht}. 


To determine $U$ experimentally the PES and IPES spectra of CeAg$_2$Ge$_2$ are shown in Figure 2(a). Here, the transitions to the lowest final state of Ce are considered which correspond to the minimum energy required to excite a Ce $4f$ electron to E$_F$ ($\Delta_-$) in case of PES and the transition which corresponds to exciting an electron from E$_F$ to the first empty $4f$ state ($\Delta_+$) in the case of IPES. $U$ has been determined from the relation $U$= $\Delta_+$-$\Delta_-$. The position of the features in PES and IPES spectra have been determined by fitting the spectra with the multiple Gaussian peaks as shown in Figure 2(b) and 2(c). Minimum numbers of peaks are considered which shows a good fitting. Background correction has been carried out by considering a Tougaard background \cite{Tougaard} for PES spectra \cite{Soma1, Soma2} and a parabolic background for the IPES spectra \cite{Pandey}. In the IPES spectra in Figure 2(c) the state which is lying near to the E$_F$ at 0.6~eV corresponds to $f^1$ configuration and the state at higher binding energy of 2.5~eV corresponds to $f^2$ configuration. The intermediate feature between $f^1$ and $f^2$ corresponds to Ce $5d$ states at 1.4~eV (Figure 2(c))\cite{Hillbrecht, Parlebas}. Ce features corresponding to $f^1$, Ce $5d$ and $f^2$ states are slightly broader in IPES spectra mainly due to the poor resolution. However IPES measurements on the other systems like LaCoO$_3$, PrCoO$_3$, $\alpha$ and $\beta$ Brass have shown a good agreement with the theory \cite{Pandey, Dhaka}. Similar kind of $f^1$, Ce $5d$ and $f^2$ features have also been seen for the $\gamma$-type Cerium and other Ce alloys like CeRh$_3$, CePd$_3$ and CeSn$_3$ \cite{Hillbrecht, Parlebas}. The energy positions of $f^0$ feature and the $f^2$ features are -1.7~eV ($\Delta_-$) and +2.5~eV ($\Delta_+$). Hence the value of $U$ determined from experiment is 4.2~eV. Similar method has been employed to determine the value of $U$ for Gd and NiO from XPS and BIS spectra \cite{Anisimov, Anisimov2}.


In order to verify the character of the features observed experimentally, we have calculated the density of states (DOS) using the FLAPW method with different $U_{eff}$. For $U_{eff}$= 0, the DOS calculation is reported in Ref.\cite{Soma1}. The broadened DOS for occupied and unoccupied part of $U_{eff}$= 0 are shown in Figure 2(a). A standard procedure has been adopted to broaden the DOS\cite{Soma, Soma1, Soma2, Sarma}. The occupied part has been broadened by adding the PDOS of Ce, Ag, and Ge as shown in Figure 3 after multiplying it with the photoionization cross-section at 80~eV\cite {Yeh}. Since the cross-section of the unoccupied part is not reported in the literature, so to broaden the unoccupied part the PDOS are added without considering the cross-section. The added DOS is then multiplied with the Fermi function at the measurement temperature and convoluted with a Voigt function. The full width at half maximum (FWHM) of the Gaussian component is taken to be the instrumental resolution in the PES and IPES measurement. The energy-dependent Lorentzian FWHM represents the life-time broadening \cite{Soma1, Sarma, Soma3}.  The inelastic background and the matrix elements are not considered. The calculated valence band for $U_{eff}$= 0 shows that the features correspond to Ce $4f$ states lie at the E$_F$ and at -1.5~eV (see Figure 2(a) and Figure 3(b)). The small changes in the calculated valence band feature are prominently visible mainly due to the high resolution (20 meV). The calculated valence band with the poor resolution (400~meV) doesn't show any difference in the spectral shape (Figure 4(c) of Ref.\cite{Soma1}). This is the main reason why the effect of the correlation has not been observed in the experimental PES spectra earlier carried out with poor resolution \cite{Soma1, Soma2}. Also, it is very clear that $U_{eff}$ =0 does not explain the IPES spectra and hence there is a need to take into account $U_{eff}$ in the DOS calculation. Hence in Figure 3, the partial densities of states (PDOS) are shown for $U_{eff}$= 0, 1.4, 2.7, 3.4, 4, 4.2 and 5.4~eV. In the experimental PES and IPES spectra of CeAg$_2$Ge$_2$ the dominant contribution mostly arises from the Ce $5d$, Ce $4f$, Ag $4d$ and Ge $4p$ states hence the PDOS of these states are shown in Figures 3(a), (b), (c) and (d) respectively.  Since $U_{eff}$ is the energy required to transfer an electron within the same $f$ band hence with increasing $U_{eff}$ changes are observed mostly in the Ce $4f$ states while the other Ce $5d$, Ag $4d$ and Ge $4p$ states remain almost similar to $U_{eff}$= 0 case in the occupied part \cite{Soma1}. However small changes in the PDOS of Ce $5d$ and Ge $4p$ with different $U_{eff}$ have been observed in the unoccupied part in Figure 3(a) and (d) respectively. This is because the Ge $4p$ and the Ce $5d$ states are partially filled states and the major unfilled states lie in the unoccupied part.  For $U_{eff}$= 0, the Ce $4f$ PDOS exists mostly in the unoccupied part with a small contribution in the occupied part. To show the contribution of the Ce $4f$ states for $U_{eff}$= 0 at the higher binding energy in the occupied part, we have multiplied Ce $4f$ PDOS by 50 and shifted it by 10. It is shown in the Figure 3 (b) by dashed line. $U_{eff}$= 0 clearly shows that the Ce $4f$ states at -1.5~eV (marked by tick in Figure 3(b)) is hybridized with the Ce $5d$, Ag $4d$ and Ge $4p$ states \cite{Soma1}.  


As $U_{eff}$ is increased in the DOS calculation (Figure 3) it is observed that the quasiparticle peak corresponding to $f^1$ state near E$_F$ vanishes and Ce $4f$ band shows a splitting in the occupied and the unoccupied part (compare Figure 3(b) with Figure 4 (a) of Ref.\cite{Soma1}). The splitting or the gap between the Ce $4f$ states in the occupied and unoccupied part increases with $U_{eff}$ as expected since the on-site Coulomb repulsion increases with the correlation energy \cite{Georges}. The disappearance of the $f^1$ peak with higher $U_{eff}$ in the occupied part of the DOS can be explained as follows: Ce $4f$ states are partially localized and partially delocalized (itinerant) type. The $f^1$ state which is very near to E$_F$ is mainly itinerant type and the $f^0$ is the localized state. The GGA calculation with $U_{eff}=0$ gives the itinerant behavior hence the $f^1$ state is visible in the DOS but the correlation energy in GGA+U causes progressive localization of the $4f$ states and leads to the change of ground state from itinerant to localized. Hence with increasing $U_{eff}$ the itinerant feature $f^1$ in the DOS calculation disappears. This has also been reported for other Ce based systems like CeRhIn$_5$ and CeIrIn$_5$ \cite{Gamza06}. We have calculated the DOS with $U_{eff}$= 4~eV and 4.2~eV near the experimental $U$ value. Comparing the experimentally determined $f^0$ (-1.7~eV) and $f^2$ (+2.5~eV) features (Figure 2(b) and 2(c)) with the theoretically determined Ce $4f$ feature for $U_{eff}$= 4.2~eV shows that the maxima related to the Ce $4f$ states appears at -2.3~eV in the occupied part and +2.4~eV in the unoccupied part of the DOS. 


To see the effect of electron correlation in the calculated DOS with different $U_{eff}$ and to compare it with the experimental PES and IPES spectra the DOS for occupied and unoccupied part has been broadened and shown in Figure 4. The interesting observation is that the calculated valence band with different $U_{eff}$ shows almost similar kind of trend with very little changes observed for the feature near -2~eV. However by taking $U_{eff}$= 4~eV or 4.2~eV in the calculation, the occupied DOS shows only one Ce $4f$ feature related to $f^0$ state where as the $f^1$ feature is missing which was found to be quite sharp as in case of $U_{eff}$= 0 (shown in Figure 2(b)) \cite{Soma1, Soma2}. This is due to the localization of the $4f$ states with the higher $U_{eff}$ as explained earlier. The above results shows that the electron-electron correlation effect on the Ce $4f$ states is not so evident in the occupied part of the electronic structure of CeAg$_2$Ge$_2$. The reason could be that the Ce $4f$ feature in the occupied part is very narrow and localized (see Figure 3). For $U_{eff}$= 4.2~eV, the FWHM of the Ce 4f feature is $\approx$100~meV. Also in the valence band the narrow Ce $4f$ states are covered by the broad features of Ce $5d$, Ag $4p$ and the Ge $4p$ states. In fact this is one of the reasons that it was difficult to infer about the presence of correlation in the system from the valence band photoemission spectra (see Fig. 4 (c) of Ref. \cite{Soma1}) obtained with the 400~meV instrumental resolution. Also Figure 3, shows that the DOS of Ce $4f$ states in the unoccupied part is $\geq$~3 times the DOS of the occupied part. This may be one of the reasons that the correlation effect is much more evident in the unoccupied part and is clearly visible in the inverse photoemission spectra. The broadened unoccupied part shows a considerable shift of Ce $4f$ feature towards the higher energy ${\it w.r.t.}$ E$_F$ with increasing $U_{eff}$.

In Figure 5 the calculated valence band for $U_{eff}$= 4.2~eV is compared with the on-resonance and off-resonance PES spectra and the calculated conduction band spectra for $U_{eff}$= 4.2~eV is compared with the IPES spectra. The arrows in Figure 5 show the position of the features obtained by fitting the experimental data (Figure 2(b) and 2(c)). The off-resonance and the IPES spectral shape shows a good matching with the calculated theoretical spectra. There are very small differences between the experiment and the theoretical calculation like the position of the features which could be related to the fact that the density-functional theory is a ground state calculation and it does not take into account the sample related effects such as the presence of antisite defects and site disorder {\it etc} \cite{Soma, Soma3}. Similar differences in the position of features in theory and experimental has also been observed for EuCu$_2$Ge$_2$ \cite{Soma}. Hence the calculated value of $U_{eff}$ ($\approx$ 4.2~eV) is found to be in excellent agreement with the estimated value of $U$ ($\approx$ 4.2~eV) from the IPES and high resolution PES experiments.

\section{Conclusions}
In conclusion, the electron-electron correlation energy for CeAg$_2$Ge$_2$ has been determined experimentally and theoretically. The effect of correlation in this system is not so evident in the occupied part due to the fact that the Ce $4f$ state is very narrow and localized and also it is covered by the broad states of Ce $5d$, Ag $4d$ and Ge $4p$. However the correlation effect is much more evident in the unoccupied part because the DOS of Ce $4f$ states in the unoccupied part is much higher as compared to the DOS of Ce $4f$ states in the occupied part. Hence the unoccupied part measured with the IPES gives a clear signature of the presence of correlation in this system. The experimentally determined $U$ from the IPES and high resolution PES shows a very good matching with the theoretically determined $U_{eff}$ for this system.

\section{Acknowledgement}
The authors wish to thank Dr. P. D. Gupta, Dr. Tapas Ganguli and Dr. G. S. Lodha for their constant encouragement and support. Dr. Aparna Chakrabarti is thanked for useful discussions and calculations. Dr. S. R. Barman and Dr. D. M. Phase are thanked for the experimental support. M.M. is thankful to Council of Scientific and Industrial Research, India for research fellowship.

\newpage
\noindent {\large Figure Captions :}\\

\noindent Figure 1. Valence band spectra of CeAg$_2$Ge$_2$ at the Ce $4d-4f$ resonance at h$\nu$ =121~eV with 400~meV resolution (at 300~K) and 20~meV resolution (at 40~K). Inset shows the on-resonance spectra recorded at h$\nu$= 121~eV and the off-resonance spectra recorded at h$\nu$= 80~eV, both with 20~meV energy resolution measured at 40~K.\\

\noindent Figure 2. (a) shows the PES spectra at on-resonance (h$\nu$= 121~eV)  and off-resonance (h$\nu$= 80~eV) and IPES spectra of CeAg$_2$Ge$_2$ compared with the calculated valence and conduction band (dashed lines) for $U_{eff}$= 0. The Ce $4f$ features marked by ticks. (b) shows the PES spectra at off-resonance and on-resonance fitted with the Gaussian peaks. (c) shows the IPES spectra fitted with the Gaussian peaks.\\

\noindent Figure 3. Partial DOS (PDOS) showing Ce $5d$ (a), Ce $4f$ (b), Ag $4d$ (c), and Ge $4p$ (d) states for CeAg$_2$Ge$_2$ calculated with different $U_{eff}$ varying from 0 to 5.4~eV. The dashed line in (b) shows the PDOS of Ce $4f$ state for the $U_{eff}$= 0, multiplied by 50 and shifted by 10 for the clarity of presentation of the Ce $4f$ state in the higher binding energy region.\\

\noindent Figure 4. Calculated valence band (VB) and conduction band (CB) with different $U_{eff}$.\\

\noindent Figure 5. Comparison between theory and experiment. Left panel shows the on-resonance and the off-resonance PES spectra compared with the calculated valence band spectra for $U_{eff}$= 4.2~eV. Right panel shows the IPES spectra compared with the calculated conduction band spectra for $U_{eff}$= 4.2~eV.\\

\end{document}